# An eco-friendly universal strategy via ribavirin to achieve highly efficient and stable perovskite solar cells

*Xianhu Wu,[1] Gaojie Xia,[1] Guanglei Cui,[1,]\* Jieyu Bi,[1] Nian Liu,[1] Jiaxin Jiang,[1] Jilong Sun,[1] Luyang Liu,[1] Ping Li,[2] Ning Lu,[1] Zewen Zuo,[1] Min Gu[3]*

X. Wu, G. Xia, Prof. G. Cui, J. Bi, N. Liu, J. Sun, J. Jiang, J. Sun, L. Liu, Prof. N. Lu, Prof. Z. Zuo

[1] *College of Physics and Electronic Information, Anhui Province Key Laboratory for Control and Applications of Optoelectronic Information Materials, Key Laboratory of Functional Molecular Solids, Anhui Normal University, Wuhu 241002, P. R. China.*

Prof. P. Li

[2] *School of Physics and Electronic Science, Zunyi Normal University, Zunyi 563006, P. R. China*

Prof. M. Gu

[3] *National Laboratory of Solid State Microstructures, Nanjing University, Nanjing 210093, P. R. China.*

\* Corresponding Author.

\* wxh@ahnu.edu.cn

\* glcui@ahnu.edu.cn

## ABSTRACT

The grain boundaries of perovskite films prepared by the solution method are highly disordered, with a large number of defects existing at the grain boundaries. These defect sites promote the decomposition of perovskite. Here, we use ribavirin obtained through bacillus subtilis fermentation to regulate the crystal growth of perovskite, inducing changes in the work function and energy level structure of perovskite, which significantly reduces the defect density. Based on density functional theory calculations, the defect formation energies of $V_I$, $V_{MA}$, $V_{Pb}$, and $Pb_I$ in perovskite are improved. This increases the open-circuit voltage of perovskite solar cells (PSCs) (ITO/PEDOT:PSS/perovskite/PCBM/BCP/Ag) from 1.077 to 1.151 V, and the PCE increases significantly from 17.05% to 19.86%. Unencapsulated PSCs were stored in the environment (humidity approximately 35±5%) for long-term stability testing. After approximately 900 hours of storage, the PCE of the ribavirin-based device retains 84.33% of its initial PCE, while the control-based device retains only 13.44% of its initial PCE. The PCE of PSCs (ITO/SnO$_2$/perovskite/Spiro-OMETAD/Ag) is increased from 20.16% to 22.14%, demonstrating the universality of this doping method. This universal doping strategy provides a new approach for improving the efficiency and stability of PSCs using green molecular doping strategies.

# 1. Introduction

Perovskite solar cells (PSCs) have achieved significant research breakthroughs over the past decade due to their outstanding photovoltaic performance, simple fabrication processes, and low production costs. Their certified power conversion efficiency (PCE) has now surpassed 26%, making them the most promising candidate for next-generation photovoltaic technologies. [1-5] However, perovskite lattices and grain boundaries contain numerous defects, such as interstitial sites and vacancies. These defects not only act as carrier recombination centers, triggering redox reactions within the perovskite layer, but also lead to irreversible degradation of the perovskite material. [6-7] This severely compromises device stability and drastically shortens the lifespan of PSCs. Under such defect conditions, commonly used methylammonium lead iodide ($MAPbI_3$) exhibits significant phase instability and is prone to decomposition. [8-9] Consequently, long-term stability remains the greatest obstacle to the commercialization of PSCs and is also a critical metric for their photovoltaic performance. Additionally, $MAPbI_3$-based PSCs exhibit notably low open-circuit voltages, likely due to the large interfacial barriers between the perovskite and charge transport layers (hole transport layer and electron transport layer), which hinder efficient charge transfer across the interfaces. [10-13]

Open-circuit voltage loss is a critical issue that cannot be overlooked in the development of PSCs. At the interfaces in contact with the perovskite layer, energy-level mismatches hinder efficient carrier transport. Additionally, defects at the perovskite film surface and grain boundaries contribute to significant non-radiative recombination and carrier loss. [14-16] To address these challenges, researchers have incorporated cesium (Cs) and formamidinium (FA) to form mixed-phase perovskites, which help mitigate perovskite decomposition. These modifications improve the stability and performance of the PSCs by reducing defects and enhancing the overall material properties.

Beyond creating new perovskite architectures, organic molecular doping has emerged as a simple yet effective approach to reduce open-circuit voltage loss, demonstrating broad application prospects. [17] Wang et al. introduced uracil as a "binder" into the perovskite, achieving dual benefits: effective defect passivation, strengthened grain boundaries, and improved stability under continuous light exposure. Moreover, uracil reinforced the interfacial adhesion between the perovskite and the $SnO_2$. [18] Jiang et al. systematically investigated the effects of astaxanthin, lycopene, and β-carotene on perovskites. They found that directional movement of delocalized electrons along the π-bridge in these molecules enhances the coordination capability of acceptors, ultimately boosting the power conversion efficiency of PSCs to 24.43%. [19] Zhi et al. employed ultraviolet (UV)-resistant molecules (cinnamate salts) to mitigate the adverse effects of UV exposure on perovskites. cinnamate salts regulates the crystallization process, resulting in larger perovskite grains and enhanced charge extraction in PSCs [20]. Sun et al. incorporated lead formate (PbFa) into the perovskite solution, effectively alleviating halogen vacancies and introducing tensile strain outside the perovskite lattice. This significantly increased the energy barriers for anion vacancy formation and migration, thereby reducing the risk of $Pb^{2+}$ leakage [21]. Wu et al. doped sodium Na-dithienylethene (NaDTE) into the perovskite, where the addition of NaDTE promoted perovskite crystal growth and increased the grain size of the perovskite film. The sulfur groups in NaDTE facilitated chemical bonding of lead cations within the perovskite. [22] The green natural antioxidant resveratrol (RES) was also introduced into the perovskite film to passivate defects. RES interacts with uncoordinated $Pb^{2+}$ in the perovskite film, achieving defect passivation. Additionally, RES increased the defect formation energy of $V_{Pb}$ and $Pb_I$ in the perovskite film. [23] These strategies demonstrate the crucial role of molecular additives in improving both the stability and the PCE of PSCs. Therefore, finding suitable molecular additives is an effective strategy to improve the performance and stability of perovskite films and interfaces.

In this study, we incorporated ribavirin into the perovskite to regulate its grain boundary and surface properties. The -C=O groups in ribavirin coordinate with under-coordinated $Pb^{2+}$, increasing the formation energy of $V_{Pb}$, $V_I$, $V_{MA}$, and $Pb_I$ in the perovskite, thereby significantly reducing defects. This approach enhanced the PCE of inverted PSCs (ITO/PEDOT:PSS/Perovskite/PCBM/BCP/Ag) from 17.05% to 19.86%, with the open-circuit voltage improving from 1.077 to 1.151 V. For n-i-p PSCs (ITO/SnO$_2$/Perovskite/Spiro-OMeTAD/Ag), the PCE increased from 20.16% to 22.14%. Long-term stability tests were conducted on unencapsulated devices stored under dark conditions (relative humidity 35±5%). After approximately 900 hours, the ribavirin-based devices retained 84.33% of their initial PCE, while the PEDOT:PSS-based devices retained only 13.44% of their initial efficiency. The ribavirin-modified devices demonstrated superior environmental stability.

## 2. Result and discussion

**Figure 1** shows the experimental details of ribavirin-doped perovskite with different molar concentrations ($1\times10^{-6}$, $2\times10^{-6}$, and $3\times10^{-6}$ mol), with the $2\times10^{-6}$ mol concentration yielding the best results. This concentration was chosen for further analysis in this study. To verify the effect of ribavirin on the morphology of the perovskite films, scanning electron microscope (SEM) measurements were conducted on both control and ribavirin-doped perovskite films. The top-view SEM images are shown in **Figures 2a and 2b**. A statistical analysis of the average grain size in **Figures 2a and 2b** was performed, with the grain size distribution shown in **Figures S1a and S1b**. It can be observed that after doping with ribavirin, the average grain size of the perovskite increased from 220 nm to 271 nm. This suggests that ribavirin promotes the crystallization of perovskite, reducing the grain boundaries, which is beneficial for improving the device performance.[24] To investigate the effect of ribavirin doping on the surface roughness (SR) of perovskite, the AFM images of the perovskite before and after ribavirin doping are shown in **Figure S2**. After ribavirin doping, the SR of the perovskite decreased from 27.9 to 23.1 nm. The smoother surface of the perovskite facilitates enhanced charge transport at the interface. The optical absorption spectra of control and ribavirin-doped perovskite are shown in **Figure 2c**. In the visible light range, the absorption intensity of the ribavirin-doped perovskite film increases, indicating an improvement in the crystallinity of the perovskite film, which contributes to enhancing the device's current. The corresponding tauc plots derived from the Optical absorption spectra are presented in

**Figure S3**. It is observed that the bandgap of the ribavirin-doped perovskite film slightly increases (from 1.606 to 1.615 eV).

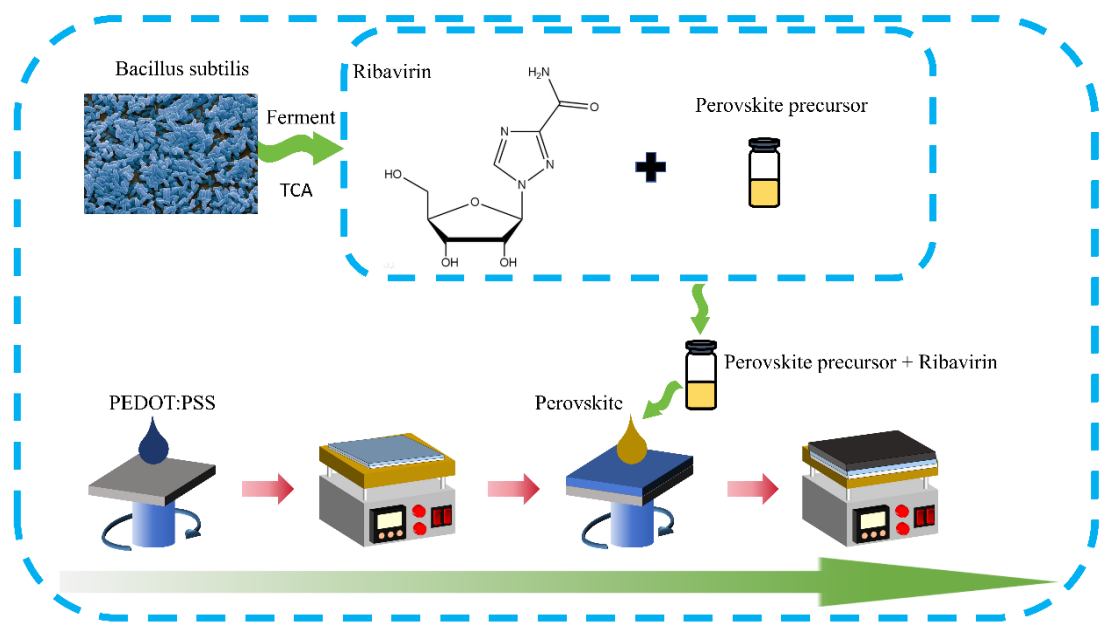

**Figure 1**. The experimental details of ribavirin-doped perovskite.

The X-ray diffraction (XRD) patterns of the control and ribavirin-doped perovskite films are shown in **Figure 2d**. The diffraction peaks corresponding to the (110) and (220) of the ribavirin-doped perovskite films exhibit significantly higher intensities compared to the control, indicating an improvement in the crystallinity of the perovskite. Additionally, the peak corresponding to $PbI_2$ is reduced in the ribavirin-doped perovskite films, suggesting that ribavirin has passivated the defects in the perovskite and enhanced its crystallization. The electrostatic potential (ESP) of ribavirin is displayed in **Figure 2e**. In the blue region, the -C=O group of ribavirin shows a high negative ESP and electron density, which further supports the notion that ribavirin plays a role in defect passivation and crystallization enhancement in the perovskite film. The oxygen atom in the -C=O group of ribavirin contains a lone pair of electrons, which allows it to function as a Lewis base. According to Lewis theory, the uncoordinated $Pb^{2+}$ in the perovskite structure can accept electrons from ribavirin molecules, leading to a Lewis acid-base coordination interaction. This interaction significantly inhibits the formation of uncoordinated $Pb^{2+}$ defects in the perovskite thin film, thus reducing the non-radiative recombination centers in the perovskite. [17,21-23,25]

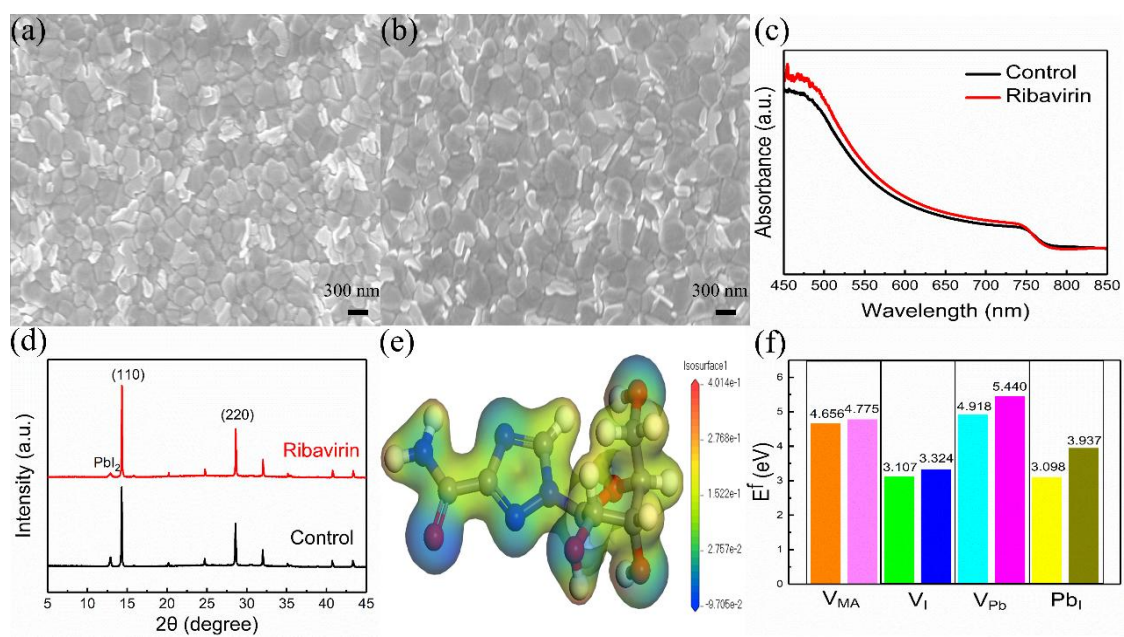

**Figure 2**. The SEM images of (a) control and (b) ribavirin-doped perovskite. (c) The optical absorption spectra of control and ribavirin-doped perovskite. (d) The XRD of control and ribavirin-doped perovskite. (e) The ESP distribution of ribavirin molecule. (f) The defect formation energy of control and ribavirin-doped perovskite.

To further investigate the effect of ribavirin on the perovskite, we performed density functional theory (DFT) calculations to study how ribavirin doping influences the formation energies of defects in the perovskite, including $V_{MA}$, $V_I$, $V_{Pb}$ and $Pb_I$. The defect formation energy calculation model is shown in **Figure S4**, and the results of the defect formation energy calculations are displayed in **Figure 2f**. Before doping with ribavirin, the defect formation energies ($V_{MA}$, $V_I$, $V_{Pb}$ and $Pb_I$) of the perovskite were 4.656, 3.107, 4.918 and 3.098 eV, respectively. After ribavirin doping, the defect formation energies for the perovskite increased to 4.775, 3.324, 5.440 and 3.937 eV, respectively. The increase in defect formation energy suggests that ribavirin doping helps suppress the formation of $V_{MA}$, $V_I$, $V_{Pb}$ and $Pb_I$. This reduction in defect formation leads to a decrease in non-radiative recombination centers within the perovskite, which benefits the improvement of the photovoltaic performance of the device. This enhanced defect passivation is key to the improvement of the device's performance after ribavirin doping.[7]

X-ray photoelectron spectroscopy (XPS) was used to measure the control and ribavirin-doped perovskite, as shown in **Figure S5.** To further investigate the interaction between ribavirin and perovskite, the XPS of the Pb 4f peak for both the control and ribavirin-doped perovskite is displayed in **Figure 3a**. In the control, the binding energies of the Pb 4f peaks are 143.33 eV and 138.51 eV, while in the ribavirin-doped perovskite, the Pb 4f peak shifts 0.30 eV towards lower binding energy. In addition, the XPS of the I 3d peaks for both the control and ribavirin-doped perovskite are shown in **Figure 3b**. The binding energies of the I $3d_{5/2}$ and I $3d_{3/2}$ peaks in the perovskite are located at 630.71 eV and 619.21 eV, respectively. After doping with ribavirin, the binding energies of the I $3d_{5/2}$ and I $3d_{3/2}$ peaks in the perovskite shift 0.25 eV towards lower binding energy. This shift towards lower binding energy indicates a strong Lewis acid-base coordination interaction between

ribavirin and the perovskite.[7,23,26]

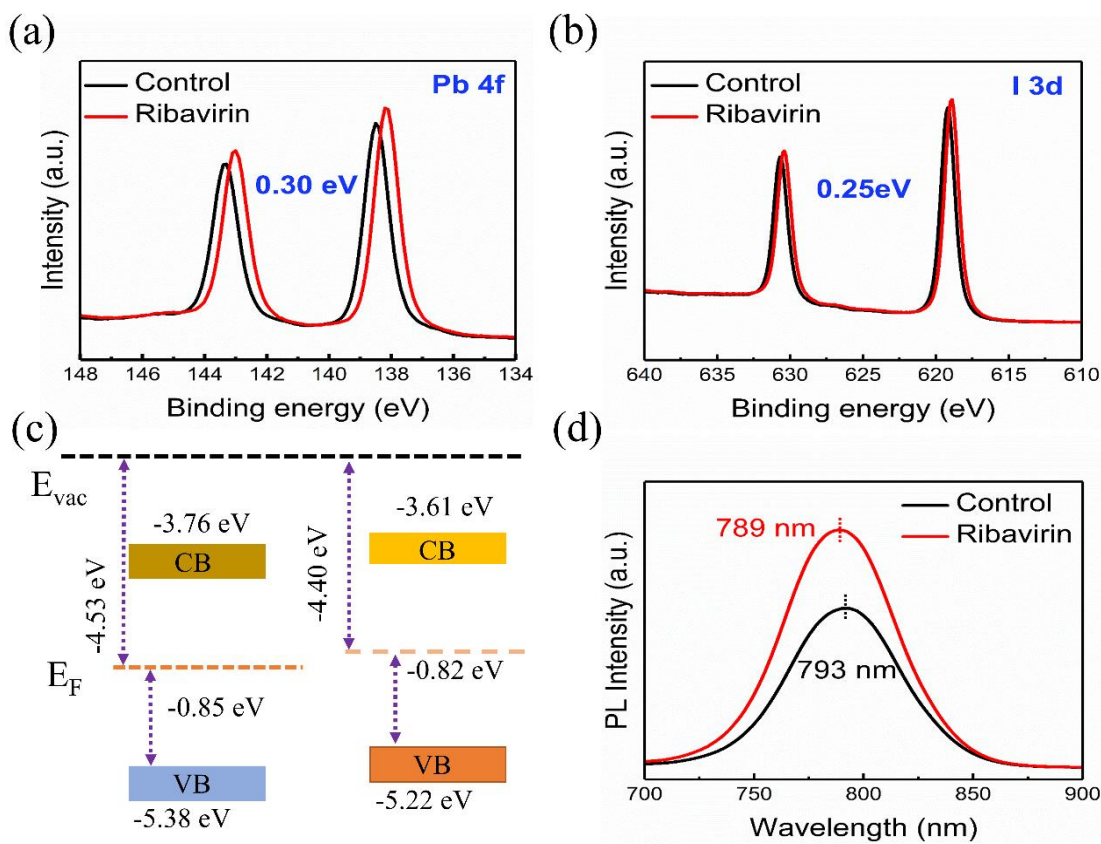

**Figure 3**. (a) The XPS spectra of Pb 4f for control and ribavirin-doped perovskite. (b) The XPS spectra of I 3d for control and ribavirin-doped perovskite. (c) Schematic energy levels of control and ribavirin-doped perovskite. (d) Steady-state PL of control and ribavirin-doped perovskite.

To further evaluate the reduction of defects, space-charge-limited current (SCLC) measurements were conducted. The SCLC of the devices based on control and ribavirin were measured and shown in **Figure S6**. The *J-V* curves include three regions[10,22]: the defect-filled region, the ohmic region, and the defect-free region. The pure hole devices show that after doping with ribavirin, the defect-filled limiting voltage $V_{TFL}$ of the device decreased from 0.956 to 0.685 V. Then, the $V_{TFL}$ can be used to calculate the defect density ($N_t$) of the sample using the following formula: $N_t = (2\varepsilon\varepsilon_0 V_{TFL})/qL^2$, where $\varepsilon$ is the relative permittivity of the perovskite, $\varepsilon_0$ is the vacuum permittivity, $q$ is the amount of charge, and $L$ is the thickness of the perovskite film (**Table S1**). The calculations based on pure hole samples show that $N_t$ decreases from $1.61 \times 10^{16}$ cm$^{-3}$ for the Control sample to $1.15 \times 10^{16}$ cm$^{-3}$ for the ribavirin-doped perovskite.

To investigate the effect of ribavirin doping on the energy level structure of perovskites, we measured the ultraviolet photoelectron spectra (UPS) of both the control and ribavirin-doped perovskites, as shown in **Figure S7**. Using the formula $E_F = E_{cut-off} - 21.22\ eV$, the work functions of the control and ribavirin-doped perovskite were calculated to be -4.53 and -4.40 eV, respectively. From the valence band spectrum in **Figure S7**, the valence band maximum (VBM) of the control and ribavirin-doped perovskite were determined to be -5.38 and -5.22 eV, respectively. By the formula $E_{VB} = E_F - E_{F,edge}$, the $E_{F,edge}$ of the control and ribavirin-doped perovskite were found to be -0.85 and -0.82 eV, respectively. Additionally, we obtained the energy level structures of the perovskite based on the optical bandgap (**Figure S3**), which is presented in **Figure 2c**. The energy level structure of the PSCs is shown in **Figure S8**, revealing that the VBM of the ribavirin-doped perovskite is closer to the hole transport layer (PEDOT:PSS). This suggests that the doping of ribavirin improves the hole transport at the interface, which is beneficial for hole extraction in the device. The steady-state photoluminescence (PL) of two samples, PEDOT:PSS/perovskite and PEDOT:PSS/ribavirin-perovskite, was measured and is shown in **Figure 2d**. Under the same conditions, the PL peak of

PEDOT:PSS/ribavirin-perovskite is lower than that of PEDOT:PSS/perovskite, which directly confirms that charge transfer at the interface is faster in the PEDOT:PSS/ribavirin-perovskite. The slight blue shift of the PL peak could be attributed to the change in the bandgap.

The performance of PSCs (ITO/PEDOT:PSS/perovskite/PCBM/BCP/Ag) based on control and ribavirin-doped perovskites is shown in **Figure 4**. **Figure 4a** presents the J-V curves of the champion devices based on control and ribavirin-doped perovskites. These curves provide insight into the electrical characteristics and performance improvements achieved by ribavirin doping, demonstrating the positive impact on device efficiency and charge transport. The device based on the control exhibited an open-circuit voltage ($V_{oc}$) of 1.077 V, a short-circuit current density ($J_{sc}$) of 20.98 mA cm$^{-2}$, a fill factor (FF) of 75.46%, and a PCE of 17.05%. When the doping amount of ribavirin was 2×10$^{-6}$ mol, there was a noticeable improvement in both $V_{oc}$ and $J_{sc}$. The device based on ribavirin showed $V_{oc}$, $J_{sc}$, FF, and PCE of 1.151 V, 22.56 mA cm$^{-2}$, 76.49%, and 19.86%, respectively. These improvements suggest that ribavirin doping enhances the overall performance of the PSCs, through better energy level alignment, reduced recombination, and optimized charge transport. The photovoltaic performance of 30 devices based on different doping amounts of ribavirin is summarized in **Table S2**, with the device based on ribavirin showing the best stability. The external quantum efficiency (EQE) shown in **Figure 4b** demonstrates an increase in the integrated $J_{sc}$ from 19.28 to 21.55 mA cm$^{-2}$, confirming the increase in current observed in the champion device in **Figure 4a**.

Dark J-V curve measurements for both the control and ribavirin-based devices are presented in **Figure S9**. Notably, compared to the control, the ribavirin-doped devices exhibit lower current in the dark state. This suggests that ribavirin doping help to reduce dark current and improve the device's performance under illumination by minimizing leakage current. This behavior can be linked to enhanced defect passivation and the optimization of charge transport in the ribavirin-doped perovskite layers. The J-V curves of devices based on control and ribavirin under different scan directions are shown in **Figures 4c and 4d**. The hysteresis index of the devices based on control and ribavirin is 0.059 and 0.001, respectively, indicating a significant suppression of hysteresis in the ribavirin-doped device. To further investigate whether the performance improvement induced by ribavirin doping is generalizable, the effect of the optimal ribavirin doping amount on device performance was measured using an n-i-p structure (ITO/SnO$_2$/perovskite/Spiro-OMETAD/Ag). The results are presented in **Figure 4e**. This additional testing on a different device architecture helps to confirm that the performance enhancement from ribavirin doping is not restricted to a particular device configuration but may be applicable to various PSCs structures. The devices based on the control have the following performance parameters: $V_{oc}$ of 1.086 V, $J_{sc}$ of 24.64 mA cm$^{-2}$, FF of 75.35%, and PCE of 20.16%. In contrast, the devices based on ribavirin exhibit improved performance with $V_{oc}$ of 1.122 V, $J_{sc}$ of 25.77 mA cm$^{-2}$, FF of 76.56%, and PCE of 22.14%. These results directly demonstrate that Ribavirin doping improves the photovoltaic performance of PSCs universally, not just limited to enhancing the performance of the p-i-n structure

(ITO/PEDOT:PSS/perovskite/PCBM/BCP/Ag). The increase in all key performance metrics ($V_{oc}$, $J_{sc}$, FF, and PCE) suggests that ribavirin doping has a broad and significant impact on device efficiency, making it a promising approach for improving PSCs performance across various device architectures.

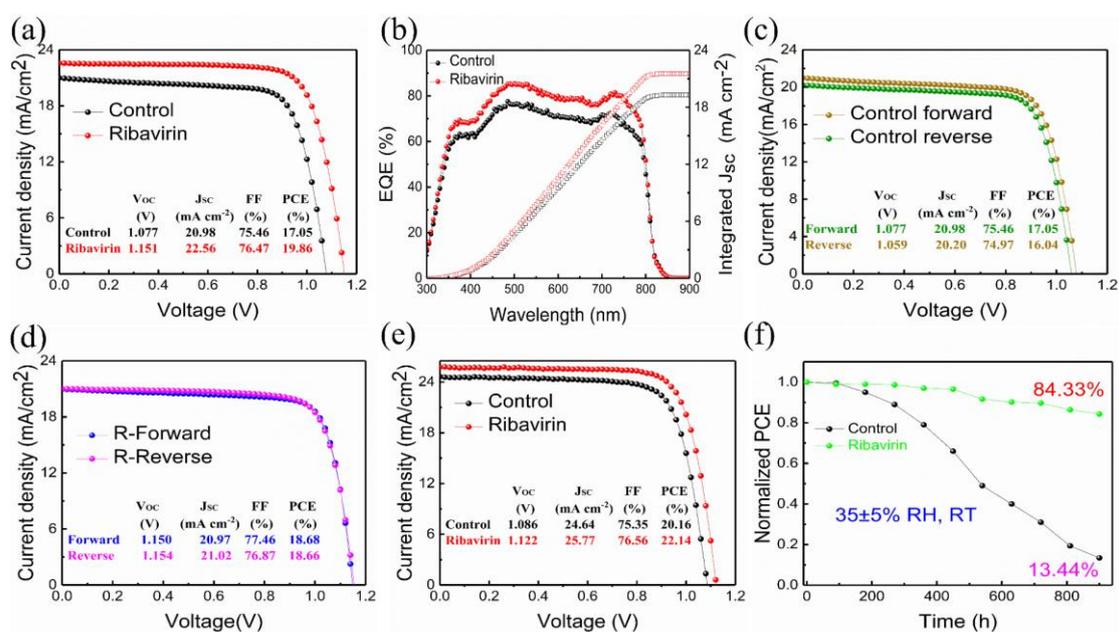

**Figure 4**. (a) The J-V curves of devices based on control and ribavirin-doped perovskite. (b) EQE spectra of devices based on control and ribavirin-doped perovskite. The J-V curves of devices based on (c) control and (d) ribavirin-doped perovskite in different scanning directions. (e) The J-V curves of devices based on control and ribavirin-doped perovskite (n-i-p structure). (f) The stability in air ambient with RH of 35 ± 5% without encapsulation of devices based on control and ribavirin-doped perovskite.

In addition to photovoltaic performance, the stability of PSCs in ambient air is also crucial.[18,21,27-28] Therefore, long-term stability tests were conducted on unencapsulated devices (p-i-n structure) under dark conditions with a humidity of approximately 35 ± 5%. As shown in **Figure 4f**, after about 900 hours of storage, the Control-based device retained only 13.44% of its initial PCE, while the ribavirin-based device retained 84.33% of its initial PCE. The superior stability of the ribavirin-based device may be attributed to a reduction in the grain boundaries and defects in the perovskite. This enhanced stability suggests that ribavirin doping not only improves device efficiency but also plays a crucial role in enhancing the durability and lifespan of PSCs in real-world conditions.

## 3. Conclusions

In summary, we proposed a universal doping strategy by using ribavirin to dope perovskite. The -C=O group in ribavirin coordinates with the uncoordinated $Pb^{2+}$, which increases the defect formation energies of $V_{Pb}$, $V_I$, $V_{MA}$, and $Pb_I$ in the perovskite. This significantly reduces the defect density, induces favorable changes in the work function, and suppresses interfacial charge recombination. As a result, the $V_{oc}$ of the device is significantly increased from 1.077 to 1.151 V, and the PCE increases dramatically from 17.05% to 19.86%. Furthermore, applying this universal strategy in the n-i-p structure, the PCE of the device was significantly enhanced from 20.16% to 22.14%. This approach provides a new method for improving the PCE and stability of PSCs using a green molecular doping strategy.

**Declaration of Competing Interest**

The authors declare that they have no known competing financial interests or personal relationships that could have appeared to influence the work reported in this paper

**Supporting Information**

Supporting Information is available from the Wiley Online Library or from the author.

**ACKNOWLEDGMENTS**

This work was supported by the National Natural Science Foundation of China through Grants 21373011 and 12264060, Anhui Provincial Natural Science Foundation (2108085MA24).

Received: ((will be filled in by the editorial staff))
Revised: ((will be filled in by the editorial staff))
Published online: ((will be filled in by the editorial staff))

**REFERENES**

Ribavirin obtained through Bacillus subtilis fermentation is used as a perovskite additive, significantly reducing the defect densities of $V_I$, $V_{MA}$, $V_{Pb}$, and $Pb_I$. This results in a substantial increase in the power conversion efficiency (PCE) of the perovskite solar cells (ITO/PEDOT:PSS/perovskite/PCBM/BCP/Ag) from 17.05% to 19.86%. The PCE of perovskite solar cells (ITO/SnO$_2$/perovskite/Spiro-OMETAD/Ag) improves from 20.16% to 22.14%, while enhancing long-term stability.

Xianhu Wu,[1] Gaojie Xia,[1] Guanglei Cui,[1,*] Jieyu Bi,[1] Nian Liu,[1] Jiaxin Jiang,[1] Jilong Sun,[1] Luyang Liu,[1] Ping Li,[2] Ning Lu,[1] Zewen Zuo,[1] Min Gu[3]

# An eco-friendly universal strategy via ribavirin to achieve highly efficient and stable perovskite solar cells

**ToC figure**

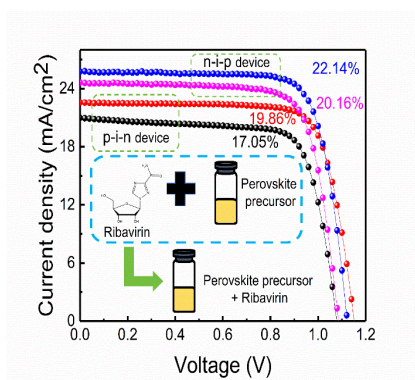